\documentclass[fleqn,twoside]{article}
\usepackage{espcrc2}
\usepackage{graphicx}
\usepackage{amsmath}
\usepackage{amssymb}
\usepackage{epsfig}

\renewcommand{\AmS}{{\protect\the\textfont2
  A\kern-.1667em\lower.5ex\hbox{M}\kern-.125emS}}

\newcommand{\www}{\mbox{
\begin{minipage}{16pt}\includegraphics[width=16pt]{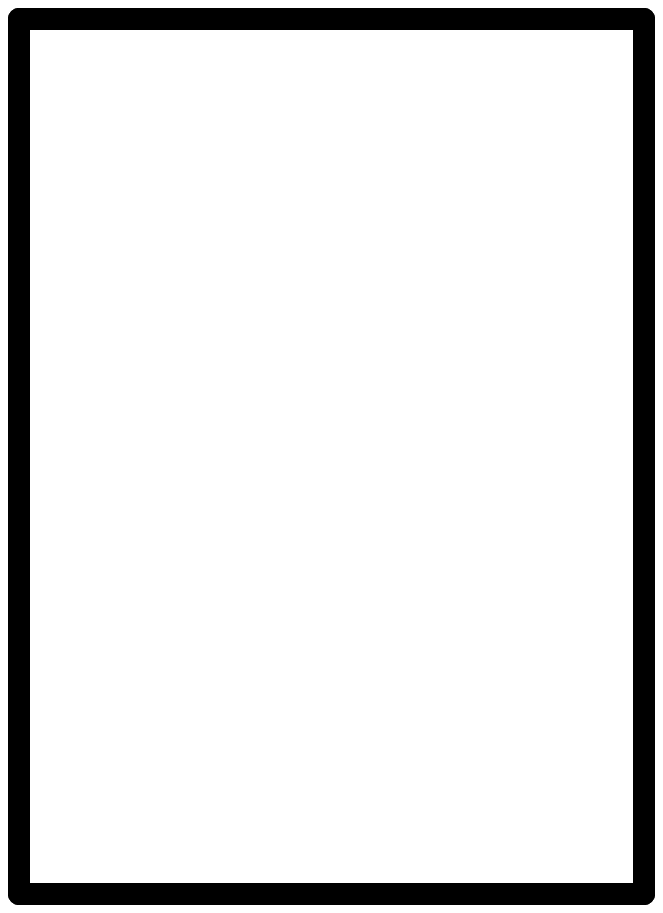}\end{minipage}}}
\newcommand{\wbbc}{\mbox{
\begin{minipage}{16pt}\includegraphics[width=16pt]{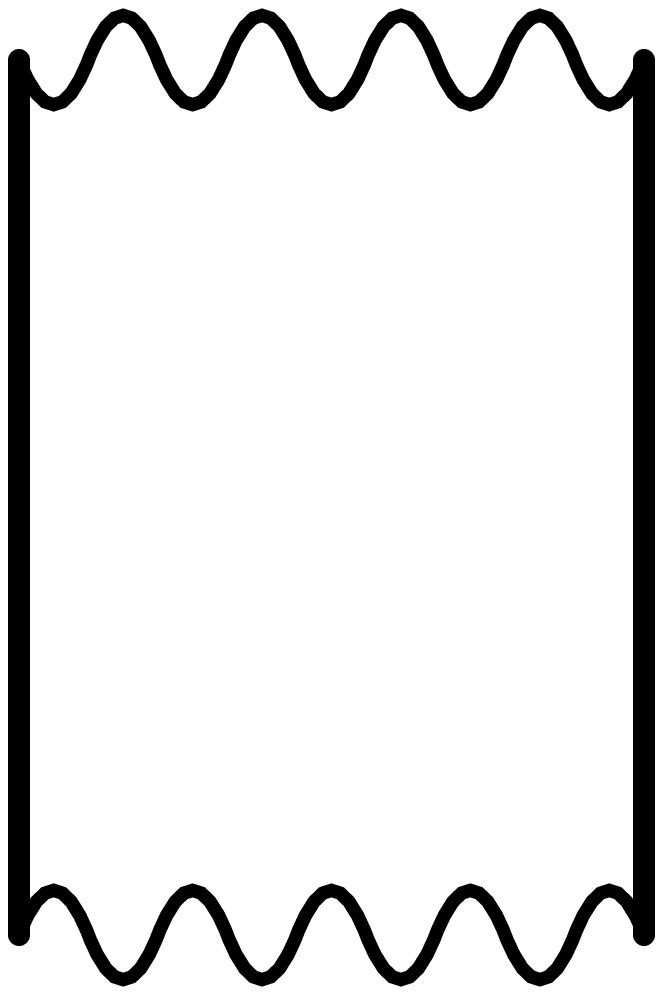}\end{minipage}}}
\newcommand{\wbbd}{\mbox{
\begin{minipage}{16pt}\includegraphics[width=16pt]{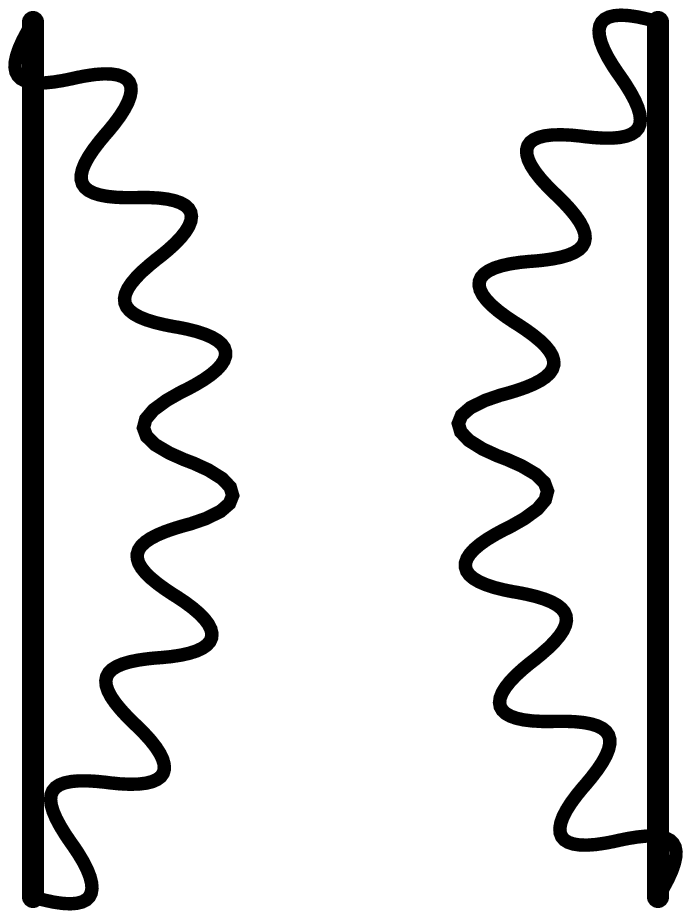}\end{minipage}}}
\newcommand{\wwb}{\mbox{
\begin{minipage}{16pt}\includegraphics[width=16pt]{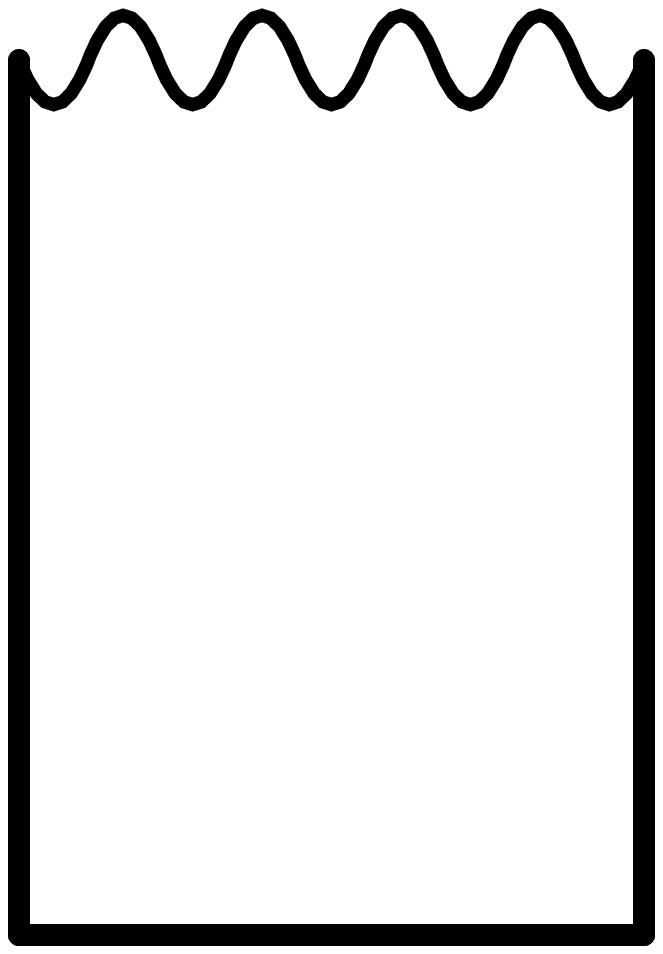}\end{minipage}}}
\newcommand{\wbw}{\mbox{
\begin{minipage}{16pt}\includegraphics[width=16pt]{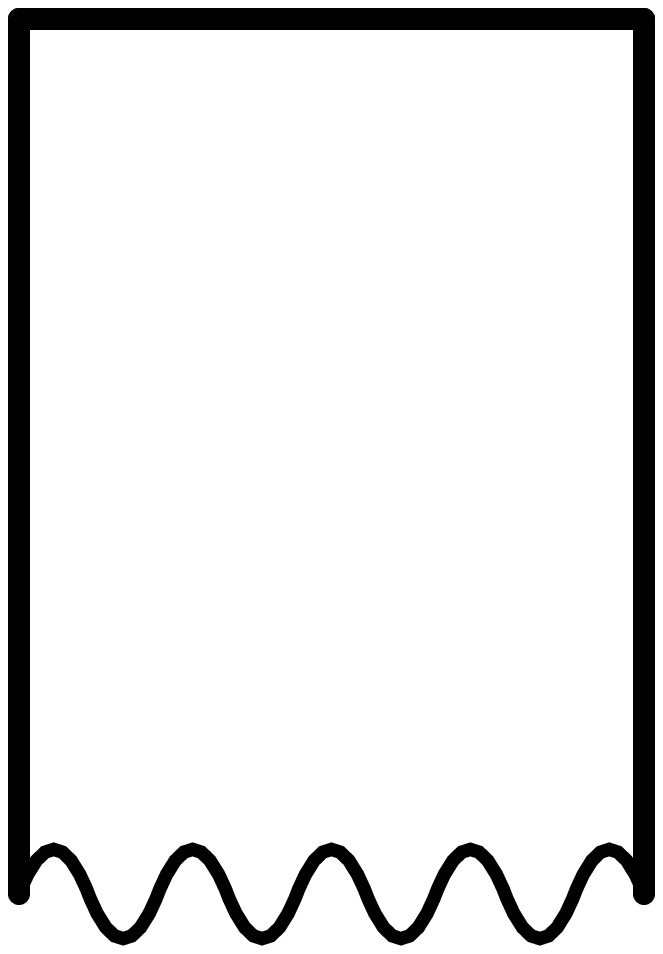}\end{minipage}}}

\title{String breaking}

\author{G.S.\ Bali\address[GLA]{Department of Physics and Astronomy,
The University
of Glasgow, Glasgow G12 8QQ, UK}\thanks{Based on talks presented
by G.S.\ Bali and K.\ Schilling.},
T.\ D\"ussel\address[JUL]{Zentralinstitut f\"ur Angewandte Mathematik,
Forschungszentrum J\"ulich, D-52425 J\"ulich, Germany},
T.\ Lippert\addressmark[JUL]\address[WUP]{Fachbereich Physik, Bergische Universit\"at Wuppertal,
D-42097 Wuppertal, Germany},
H.\ Neff\address{CCS,
Chemistry Department, University College London,
20 Gordon Street, London WC1H 0AJ, UK},
Z.\ Prkacin\addressmark[JUL] and
K.\ Schilling\addressmark[WUP]}
       
\begin{document}

\begin{abstract}
We numerically investigate the transition
of the static quark-antiquark string into a static-light
meson-antimeson system.
Improving noise reduction techniques, 
we are able to resolve
the signature of string breaking dynamics 
for $n_f=2$ lattice QCD at zero temperature.
We discuss the lattice techniques used and present results on
energy levels and mixing angle of the static
two-state system.
We visualize the
action density distribution in the region of string breaking
as a function of the static colour source-antisource separation.
The results can be related to
properties of quarkonium systems.
\end{abstract}

\maketitle

\section{INTRODUCTION}
The breaking of the colour-electric string between two
static sources is a prime example of a
strong decay in QCD~\cite{Michael:2005kw}. Recently, we
reported on an investigation of this two state
system~\cite{Bali:2004pb,Bali:2005fu,Prkacin:2005dc},
with a wave function
$|Q\rangle$ created by a $\overline{Q}Q$ operator
and a wave function $|B\rangle$ created by a four-quark $B\overline{B}$
operator, where $B=\overline{Q}q$. $Q$ denotes a static
source and $q$ is a light quark.

We determined the energy levels
$E_1(r)-2m_B$ and $E_2(r)-2m_B$ of the
two physical eigenstates $|1\rangle$ and $|2\rangle$ which
we decomposed into the components,
\begin{eqnarray}
\label{eq:1}
|1\rangle&=&\cos{\theta}\,|Q\rangle+\sin{\theta}\,|B\rangle\\
|2\rangle&=&-\sin{\theta}\,|Q\rangle+\cos{\theta}\,|B\rangle.
\label{eq:2}
\end{eqnarray}
We characterise string breaking
by the distance scale $r_c$ at which $\Delta E = E_2-E_1$
is minimized and by the
energy gap $\Delta E_c=\Delta E(r_c)$. While these energy levels
and $r_c$ are first principles
QCD predictions, the mixing angle $\theta$ is (slightly)
model dependent: within each (Fock) sector there are further
radial and gluonic excitations and we truncated the basis
after the four quark operator.

In order to obtain dynamical information on the string breaking
mechanism, we are studying the spatial
energy and action density distributions within
the two state system. In doing so one can address
questions about the localisation of the light $q\bar{q}$ pair
that is created when $r$ is increased beyond $r_c$.
The energy density will decrease fastest
in those places where $q\bar{q}$ creation is most likely.
Perturbation theory suggests that
light pair creation close to one of the static sources
is favoured by the Coulomb energy gain while aesthetic arguments
might suggest a symmetric situation with $q\bar{q}$
dominantly being created near the centre. 
First results on this investigation are presented here.

\section{PARAMETERS AND METHODS}
We use $n_f=2$ Wilson fermions at a quark mass
slightly smaller
than the physical strange quark and a lattice spacing
$a=0.166(2)\,r_0\approx 0.083(1)$~fm and
find~\cite{Bali:2005fu}, $r_c=2.5(3)\,r_0\approx 1.25(1)\,\mbox{fm}$
and $\Delta E_c\approx 51(3)$~MeV, where the errors do
not include the phenomenological uncertainty of assigning a physical
scale to $r_0\approx 0.5$~fm.
Using data on the $\overline{Q}Q$ potential and the
static-light meson mass $m_B$, obtained at different quark masses,
we determine the real world estimate, $r_c= 1.13(10)(10)$~fm,
where the errors reflect all systematics. An extrapolation
of $\Delta E_c$ however is impossible,
without additional
simulations at lighter quark masses.

These results became possible by combining a variety of improvement
techniques: the necessary all-to-all light quark propagators were
calculated from the lowest eigenmodes of the
Wilson-Dirac operator, multiplied by $\gamma_5$, after a
variance reduced stochastic estimator correction step. The signal
was improved by employing a fat link static
action.
Many off-axis distances were implemented to allow
for a fine spatial resolution of the string breaking region.
For details see Ref.~\cite{Bali:2005fu}.

\begin{figure}[th]
\epsfig{file=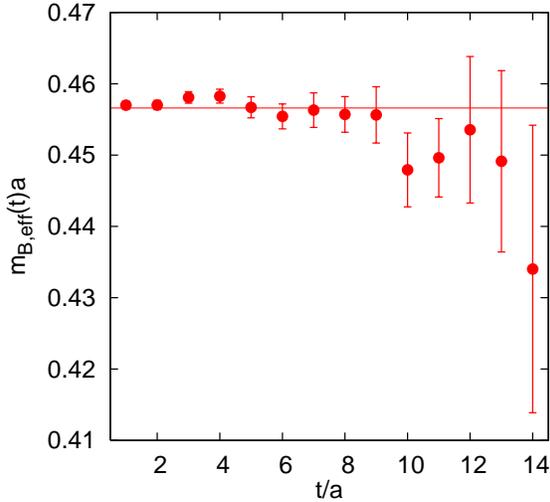,width=0.45\textwidth}\\[-1.2cm]
\caption{Static-light effective masses, $am_{B,\mbox{eff}}=\ln
[C_B(t)/C_B(t+a)]$, where $C_B(t)$ is the static-light mesonic correlation
function.}
\label{fig:effmass}
\end{figure}
Great care was also spent on optimizing
the overlap with the ground states, within
the $|Q\rangle$ and $|B\rangle$ sectors,
using combinations of APE~\cite{Albanese:1987ds,Teper:1987wt}
and Wuppertal~\cite{Gusken:1989ad} smearing.
Our APE smearing consists of the replacement,
\begin{equation}
\label{eq:smear}
U_{x,i}^{(n+1)}= P_{SU(3)}
\left(\alpha\,U_{x,i}^{(n)}+\Sigma_{x,i}^{(n)}\right).
\end{equation}
$P_{SU(3)}$ denotes a projection operator, back onto the $SU(3)$ group,
and 
\begin{equation}
\Sigma_{x,i}^{(n)}=
\sum_{|j|\neq i}
U_{x,j}^{(n)}U^{(n)}_{x+a\hat{\boldsymbol{\jmath}},i}U^{(n)\dagger}_{x+a\hat{\boldsymbol{\imath}},j},
\end{equation}
$i\in\{1,2,3\}, j\in\{\pm 1,\pm 2,\pm 3\}$,
is the spatial staple-sum, surrounding $U_{x,i}^{(n)}$.
We choose $\alpha=2.0$ and define our APE smeared links,
$\widetilde{U}_{x,i}
=U_{x,i}^{(50)}$. For the projection we use,
\begin{eqnarray}
P_{SU(3)}(A)&=&A'\det(A')^{-1/3}\in SU(3),\nonumber\\
A'&=&\frac{A}{\sqrt{A^{\dagger}A}}\in U(3).
\end{eqnarray}

The spatial transporters within our $\overline{Q}Q$ states are
products of APE smeared links, taken along the shortest
lattice distance between the two sources.
The APE smeared links
are also employed for the parallel transport within
the Wuppertal smearing of light quark sources,
used to improve the
static-light meson operators:
\begin{equation}
\label{eq:wuppertal}
\phi^{(n+1)}_x=\frac{1}{1+6\delta}\left(\phi^{(n)}_x+
\delta\sum_{j=\pm 1}^{\pm 3}\widetilde{U}_{x,j}\phi^{(n)}_{x+a\hat{\boldsymbol{\jmath}}}\right).
\end{equation}
We set $\delta=4$ and take
the linear combination $\phi^{(20)}-6.6323\phi^{(40)}+7.2604\phi^{(50)}$ as
our smearing function.
Best results are
obtained by using smeared-local quark propagators.
The quality of the overlap with the static-light mesonic ground state
is visualized in the effective mass plot Figure~\ref{fig:effmass}. Note that
we display the effective masses up to physical distances $t>1.2$~fm.

\begin{figure}[th]
\epsfig{file=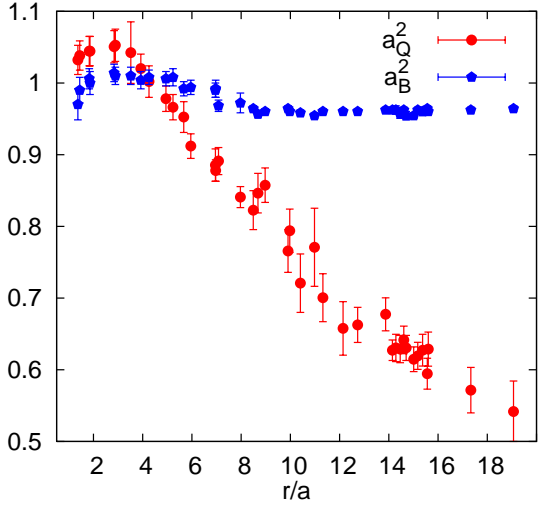,width=0.44\textwidth}\\[-1.2cm]
\caption {The overlaps of our test wave functions with the
$|Q\rangle$ and $|B\rangle$ Fock states. String breaking takes
place around $r_c\approx 15\,a$.}
\label{fig:overlap}
\end{figure}
We were able to achieve
values of 
$|a_Q|^2=|\langle \Psi_Q|Q\rangle|^2=0.62(2)$ and
$|a_B|^2=|\langle \Psi_B|B\rangle|^2=0.96(1)$ at
$r\approx r_c$ for the overlaps of
our test wave functions $|\Psi_X\rangle$
with the respective states
on the right hand sides of Eqs.~(\ref{eq:1}) and (\ref{eq:2}).
We display these results in Figure~\ref{fig:overlap}.
The almost optimal value $|a_B|\approx 1$ was essential
to allow $E_1$, $E_2$ and $\theta$ to be fitted from
correlation matrix data,
\begin{eqnarray}
C(t)&=&\left(\begin{array}{cc}C_{QQ}(t)&C_{QB}(t)\\
C_{BQ}(t)&C_{BB}(t)\end{array}\right)\label{eq:sb}
\\
&=&\left(\begin{array}{cc}
\www& \sqrt{n_f}\wwb\\&\\
\sqrt{n_f}\wbw&-n_f\wbbc+\wbbd\end{array}\right),\nonumber
\end{eqnarray}
obtained at moderate
Euclidean times:
$t\geq 2a$ for $C_{BB}$ and
$t\geq 4a$ for the remaining 2 matrix elements $C_{QQ}$ and
$C_{QB}=C_{BQ}$ at $r\approx r_c$.

We follow Ref.~\cite{Bali:1994de} and
define action and energy density distributions,
\begin{eqnarray}\label{eq:sig}
\sigma_n({\mathbf x})=\frac12\left[{\mathcal E}_n({\mathbf x})+
{\mathcal B}_n({\mathbf x})\right],\\
\epsilon_n({\mathbf x})=\frac12\left[{\mathcal E}_n({\mathbf x})-{\mathcal B}_n({\mathbf x})\right],
\label{eq:eps}
\end{eqnarray}
where
\begin{eqnarray}
\label{eq:cal}
{\mathcal A}_n({\mathbf x})&=&\langle n|A^2({\mathbf x})|n\rangle
-\langle A^2\rangle\\\nonumber
&=&\lim_{t\rightarrow\infty}\frac
{\langle \Phi_n(t)|A^2({\mathbf x,t/2})|\Phi_n(0)\rangle}
{\langle \Phi_n(t)|\Phi_n(0)\rangle}-\langle A^2\rangle
\end{eqnarray}
$\Phi_n$ are our approximations to the creation operators
of the states $|n\rangle$, obtained from the
diagonalisation of $C(t)$.
We have suppressed the distance $r$ from the above formulae
and $n=1$ denotes the ground state (dominantly $\overline{Q}Q$
at $r<r_c$) and $n=2$ the excitation (dominantly $B\overline{B}$
at $r<r_c$).
Electric and magnetic fields are calculated from the
plaquette,
\begin{equation}
E^2\left(x+\frac{a}{2}\hat{\mathbf 4}\right)=\frac{\beta}{a^4}\sum_{i=1}^3
\left[\overline{U}_{x,i4}+\overline{U}_{x-a\hat{\boldsymbol{\imath}},i4}\right],
\end{equation}
\begin{eqnarray}
B^2(x)=\frac{\beta}{2a^4}\sum_{i=1}^3\!\!\!\!\!\!\!\!
&&\left[\overline{U}_{x,ij}(x)+\overline{U}_{x-a\hat{\boldsymbol{\imath}},ij}\right.\\\nonumber
&&\left.+\overline{U}_{x-a\hat{\boldsymbol{\jmath}},ij}
+\overline{U}_{x-a(\hat{\boldsymbol{\imath}}+\hat{\boldsymbol{\jmath}}),ij}
\right],
\end{eqnarray}
where $j=\mbox{mod}(i,3)+1$ and,
\begin{equation}
\label{eq:nn}
\overline{U}_{x,\mu\nu}=\frac{z_0}{3}\mbox{tr}\,\left(\overline{U}_{x,\mu}
\overline{U}_{x+a\hat{\boldsymbol{\mu}},\nu}
\overline{U}^{\dagger}_{x+a\hat{\boldsymbol{\nu}},\mu}
\overline{U}^{\dagger}_{x,\nu}\right).
\end{equation}
We implement two different operators with the same
continuum limits: in one case we identify
$\overline{U}_{x,\mu}$ with the link
$U_{x,\mu}$ connecting $x$ with $x+a\hat{\boldsymbol{\mu}}$.
Additionally, we used smeared operators,
\begin{equation}
\overline{U}_{x,\mu}=
P_{SU(3)}\left(\gamma\,U_{x,\mu}+\Sigma'_{x,\mu}\right),
\end{equation}
where $\gamma = 0.4$ and $\Sigma'_{x,\mu}$ is the sum over all six staples
enclosing $U_{x,\mu}$,
in the three
forward and in the 
three backward directions. The $\gamma$-value was tuned to maximize
the average plaquette, calculated from the smeared links.
For the un-smeared plaquette
$z_0=1$ in Eq.~(\ref{eq:nn}) while for smeared plaquettes
$z_0=1+O(\alpha_s)$ is adjusted such that the vacuum expectation value
of the average plaquette remains unchanged.

\begin{figure}[th]
\epsfig{file=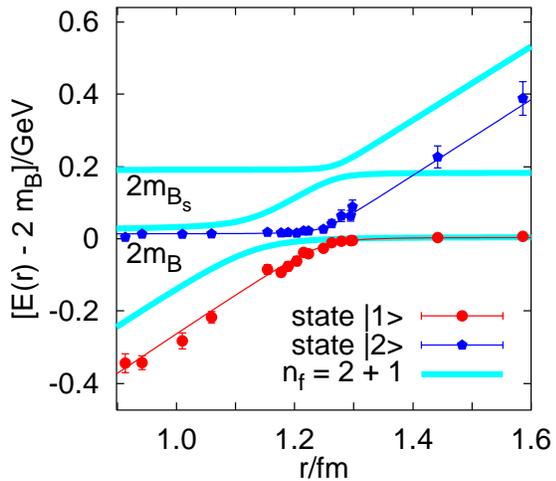,width=0.45\textwidth}\\[-1.2cm]
\caption{The energy levels
in physical units for $n_f=2$. The bands
represent the expected $n_f=2+1$ scenario.}
\label{fig:pot}
\end{figure}
\begin{figure}[th]
\epsfig{file=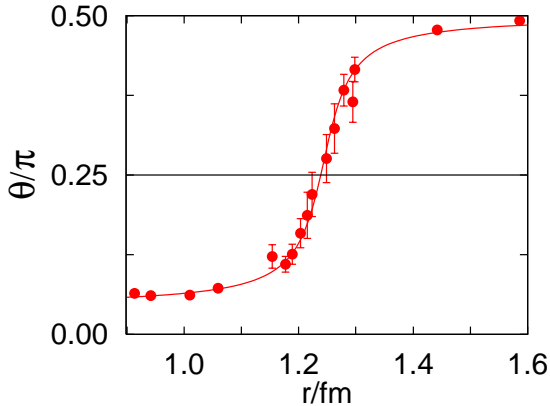,width=0.45\textwidth}\\[-1.2cm]
\caption{The mixing angle $\theta$.}
\label{fig:angle}
\end{figure}
The plaquette smearing enhances the signal/noise
ratio. Due to this smearing and the fat link static
action used, the peaks of the distributions around the source
positions (that will diverge in the continuum limit)
are less singular than in previous studies of
$SU(2)$ gauge theory at similar lattice
spacings~\cite{Bali:1994de}.
In the continuum limit the results from smeared
and un-smeared plaquette probes will coincide, away from
these self energy peaks. The draw back of plaquette smearing
is that exact reflection
positivity is violated. However, our wave functions are
sufficiently optimized to compensate for this.

We insert the $E^2({\mathbf x},t)$ and $B^2({\mathbf x},t)$ operators
at position $t/2$ into the correlation matrix $C(t)$, Eq.~(\ref{eq:sb}).
For even$|$odd $t/a$-values
we average $E^2|B^2$ over the two adjacent time slices, respectively.
Using the fitted ground state overlap ratio $a_Q/a_B$ 
and the mixing angle $\theta$ as inputs, we
calculate the action and energy density
distributions Eqs.~(\ref{eq:sig}) and (\ref{eq:eps})
in the limit of large $t$ via Eq.~(\ref{eq:cal})
from the measured matrix elements. The distributions agree within errors
within the time range $3a\leq t \leq 6a$. The results presented here
are based on our $t=4a$ analysis.

\section{RESULTS}
To set the stage,
we display the main results of
Ref.~\cite{Bali:2005fu} in Figures~\ref{fig:pot}
and~\ref{fig:angle}.
In the first figure we also speculate about the
scenario in the real world with possible decays into
$B\overline{B}$ as well as into $B_s\overline{B}_s$.
For our parameter
settings and $n_f=2$
string breaking occurs at a distance $r_c\approx 1.25$~fm.
In Figure~\ref{fig:angle} we show the mixing angle as a function of the
distance. The $B\overline{B}$ content of the ground state is given
by $\sin\theta$. Within our statistical errors $\theta$ reaches
$\pi/4$ at $r=r_c$. Remarkably, there is a significant four quark
component in the ground state at $r<r_c$ while for $r>r_c$ 
the limit $\theta\rightarrow\pi/2$ is rapidly approached.

\begin{figure}[th]
\epsfig{file=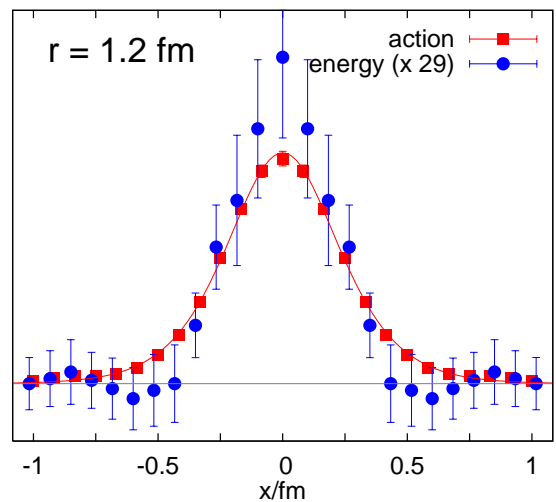,width=0.45\textwidth}\\[-1.2cm]
\caption {The transverse profile in the centre of the flux tube.}
\label{fig:transverse}
\end{figure}
Our basis vectors $|Q\rangle$ and $|B\rangle$ are no eigenstates of the
system. The transition rate, $g\propto dC_{QB}/dt$, can be related
to energy gap and mixing angle: $g(r)\approx\Delta E(r)\sin(\theta)\cos(\theta)$.
This means that $\Delta E_c\approx 2g(r)$.
With a transition rate of only about 25~MeV in the string breaking region
and even smaller $g(r)$-values at $r>r_c$, a detection of the ground state
contribution to the standard Wilson loop at $r>r_c$, where
$|1\rangle$ contains little $|Q\rangle$ admixture, is hopeless:
one would have to 
resolve the correlation function at times of order $1/g\approx 8$~fm!
For a complementary view on the problem, in the context of
the breaking of higher representation strings in pure gauge theories,
see e.g.~Ref.~\cite{Gliozzi:2004cs}.

\begin{figure*}[th]
\epsfig{file=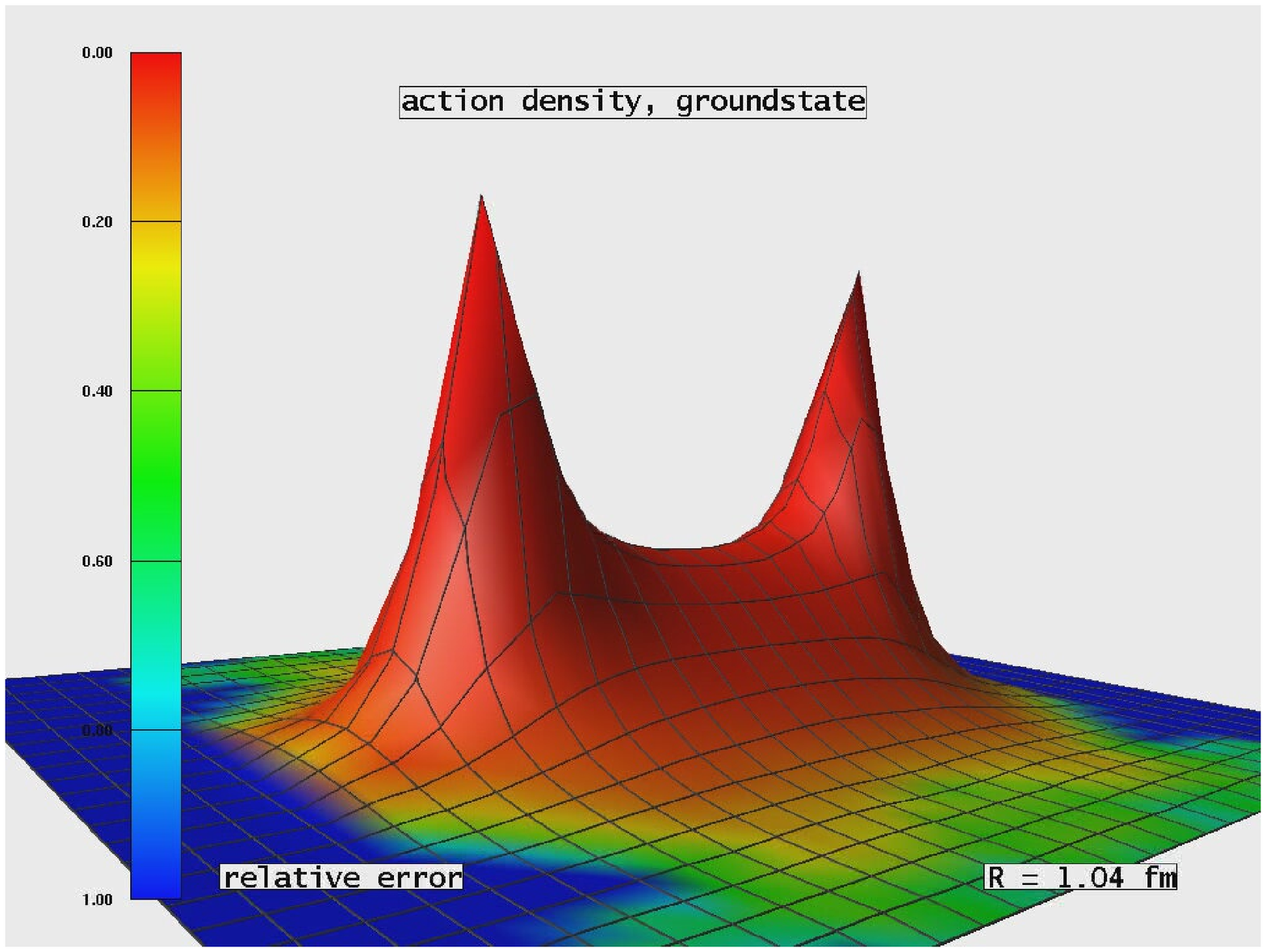,width=0.49\textwidth}~~
\epsfig{file=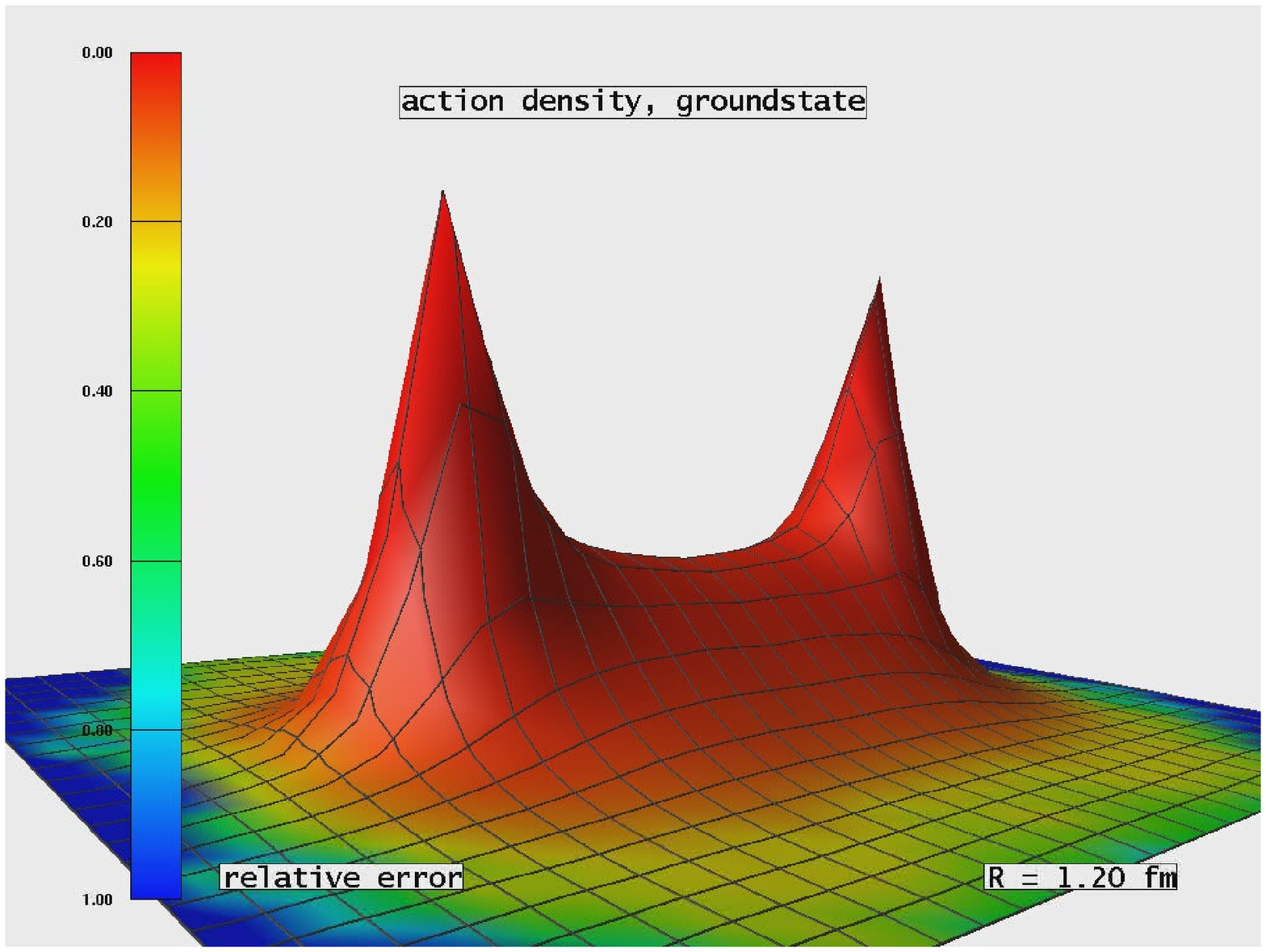,width=0.49\textwidth}\\
\epsfig{file=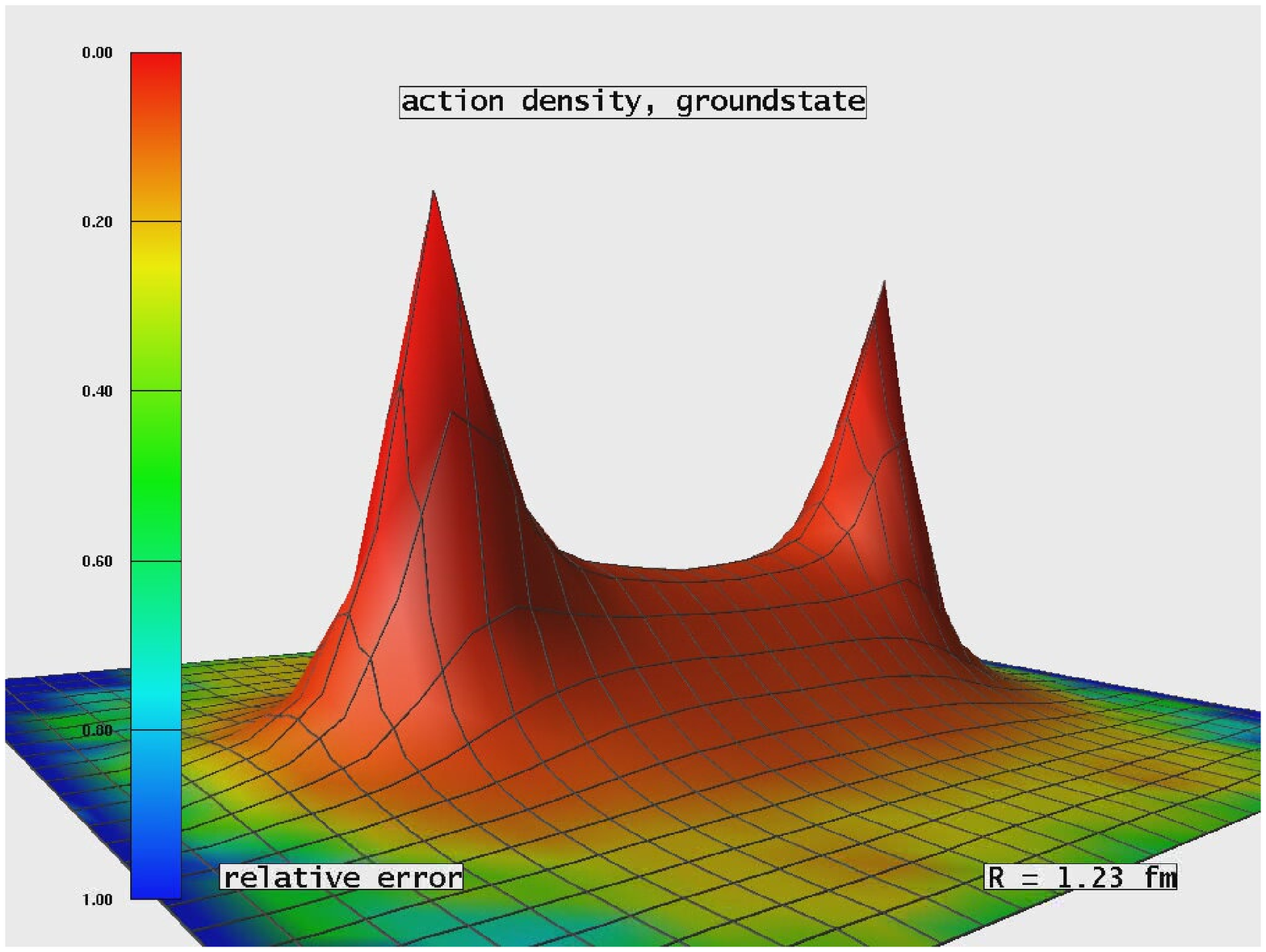,width=0.49\textwidth}~~
\epsfig{file=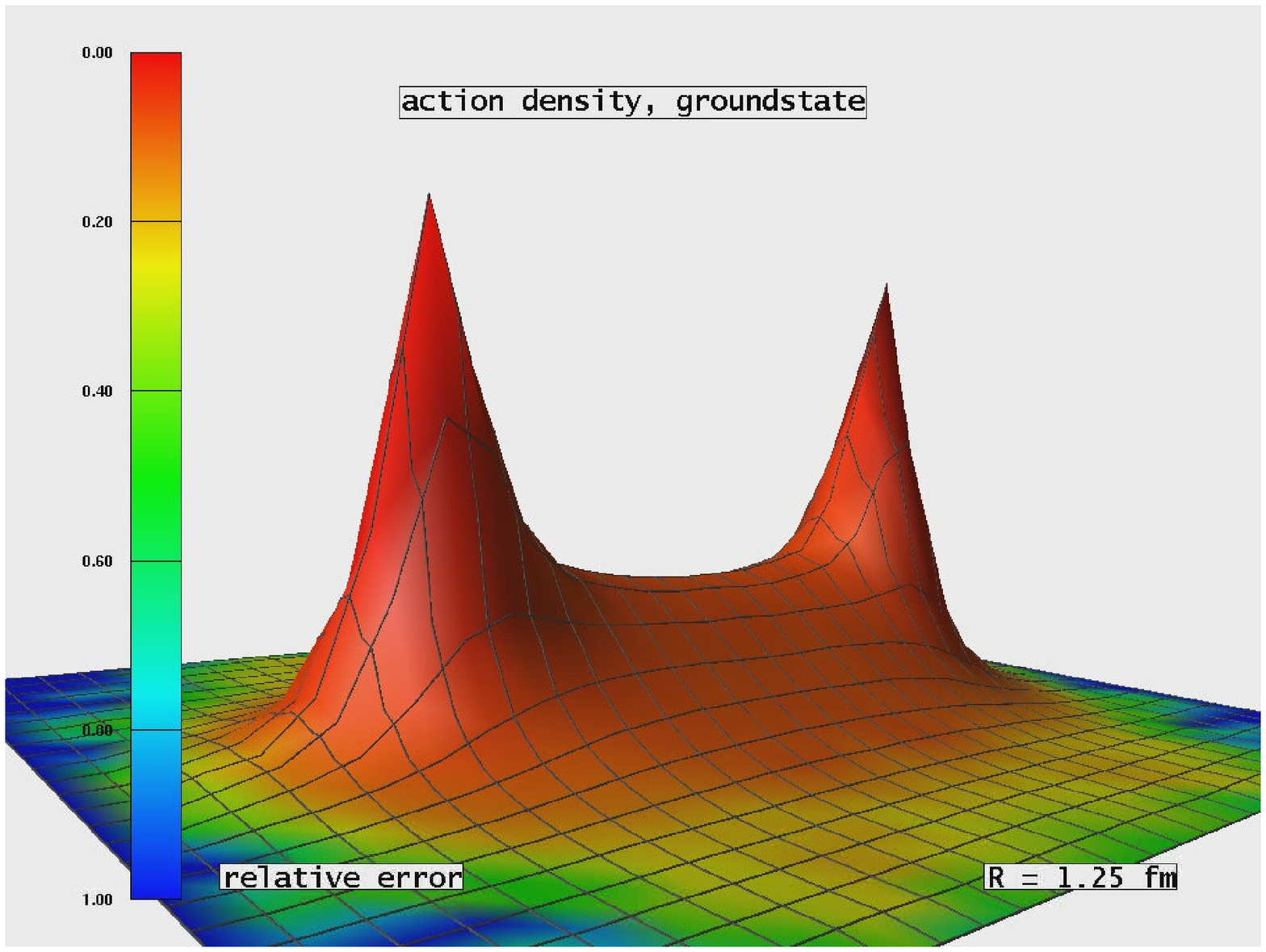,width=0.49\textwidth}\\[-1.2cm]
\caption{Ground state action density distribution at
$r\leq r_c$.}
\label{fig:frame}
\end{figure*}
\begin{figure*}[th]
\epsfig{file=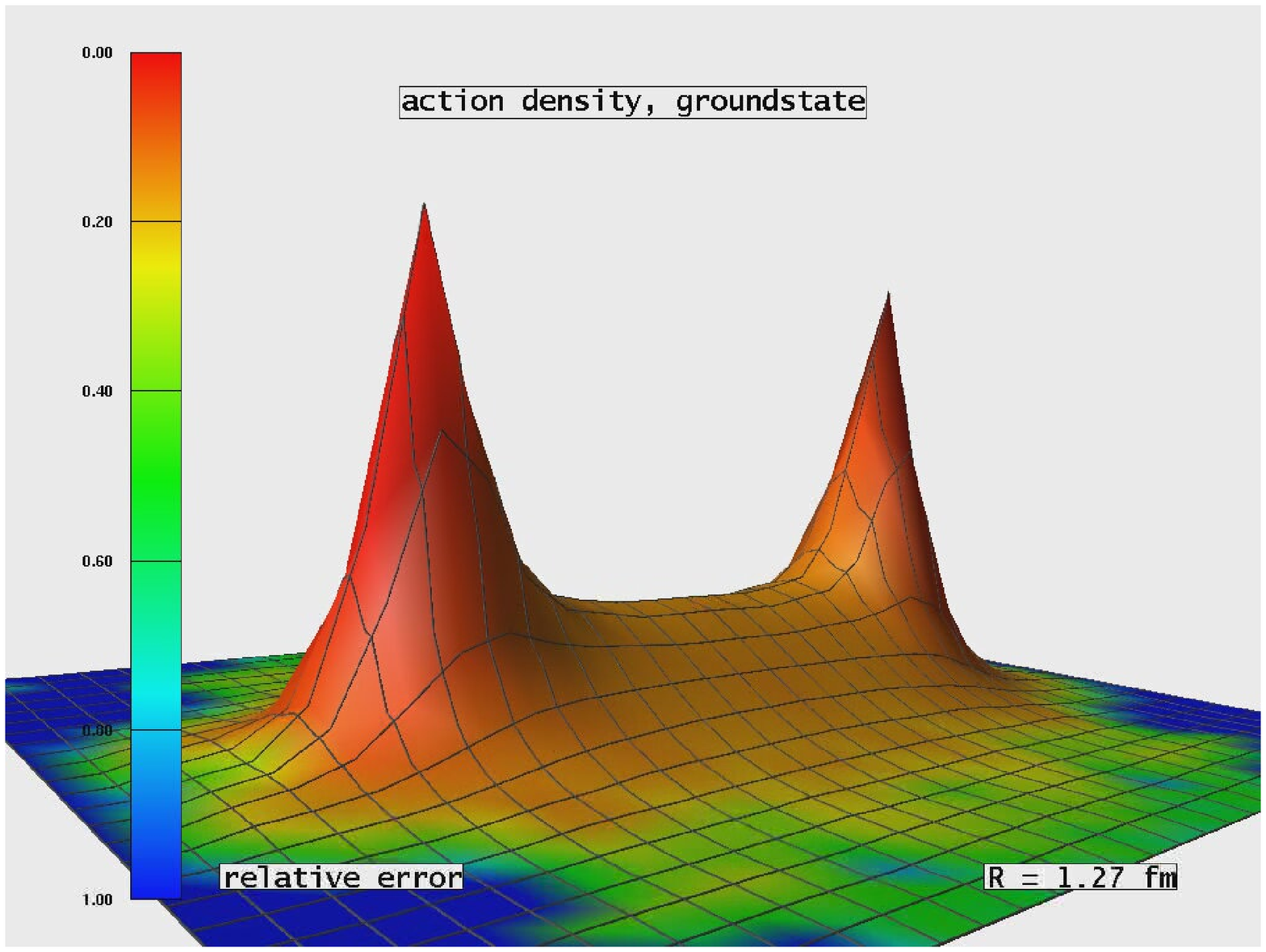,width=0.49\textwidth}~~
\epsfig{file=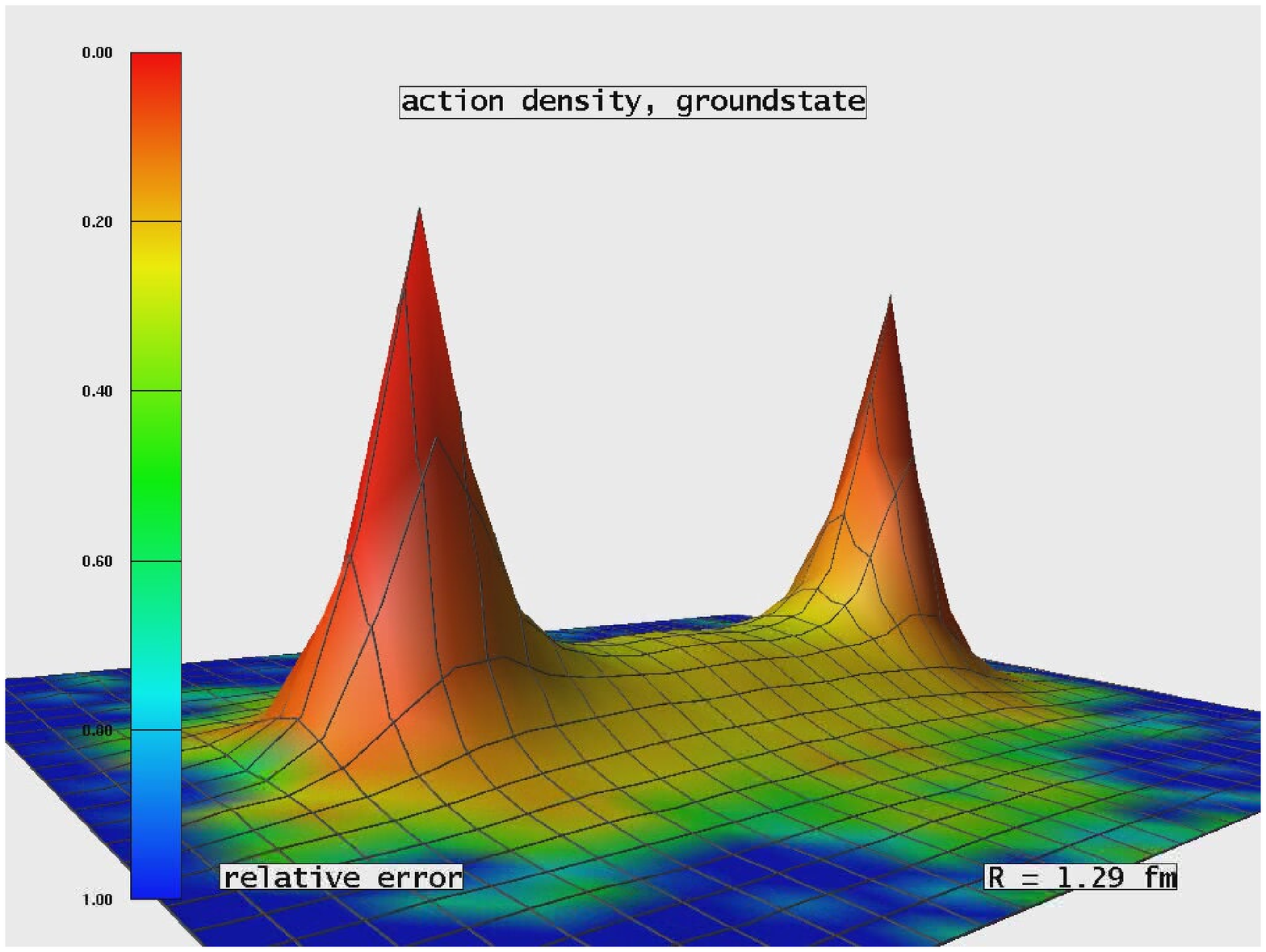,width=0.49\textwidth}\\
\epsfig{file=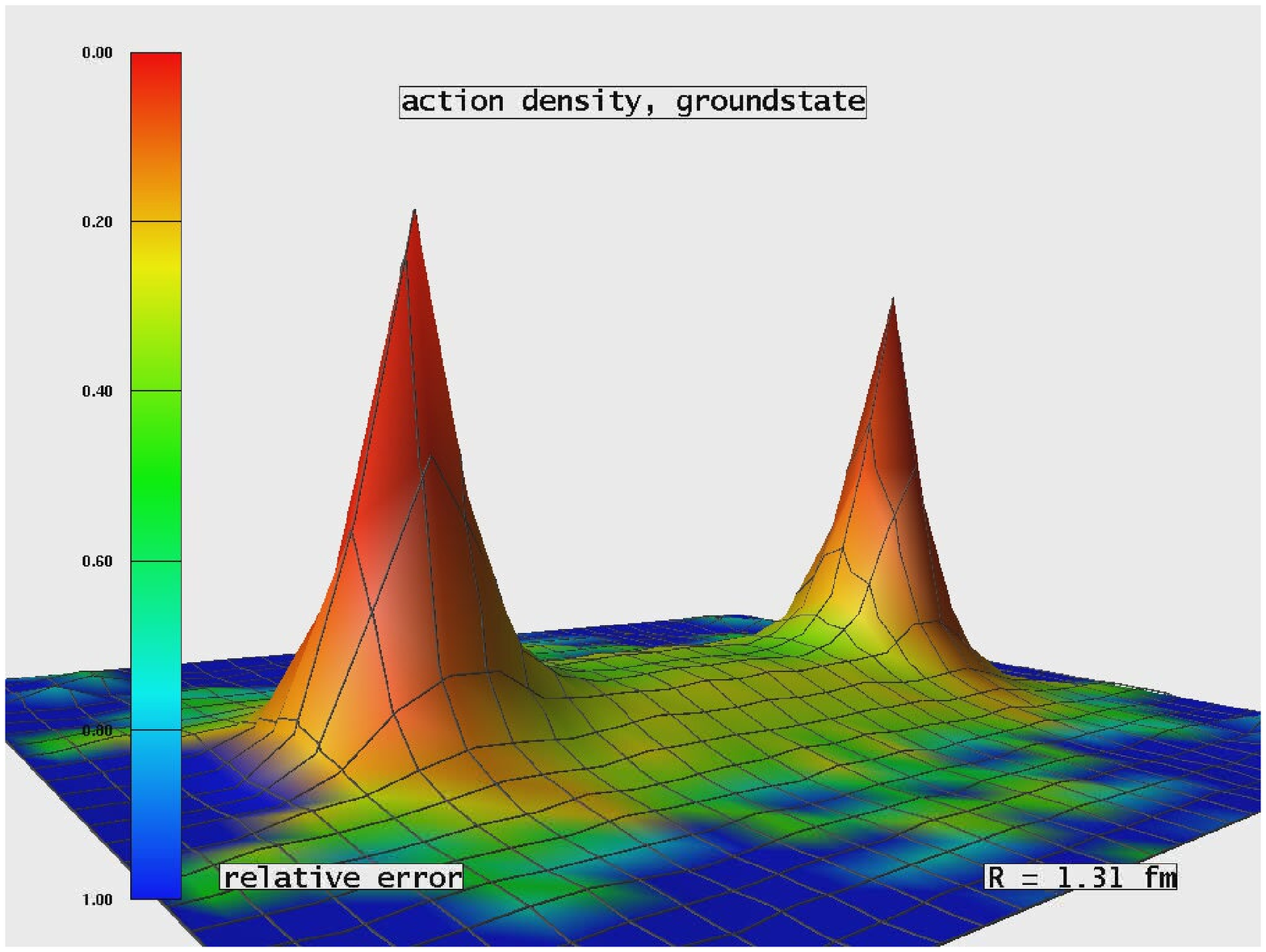,width=0.49\textwidth}~~
\epsfig{file=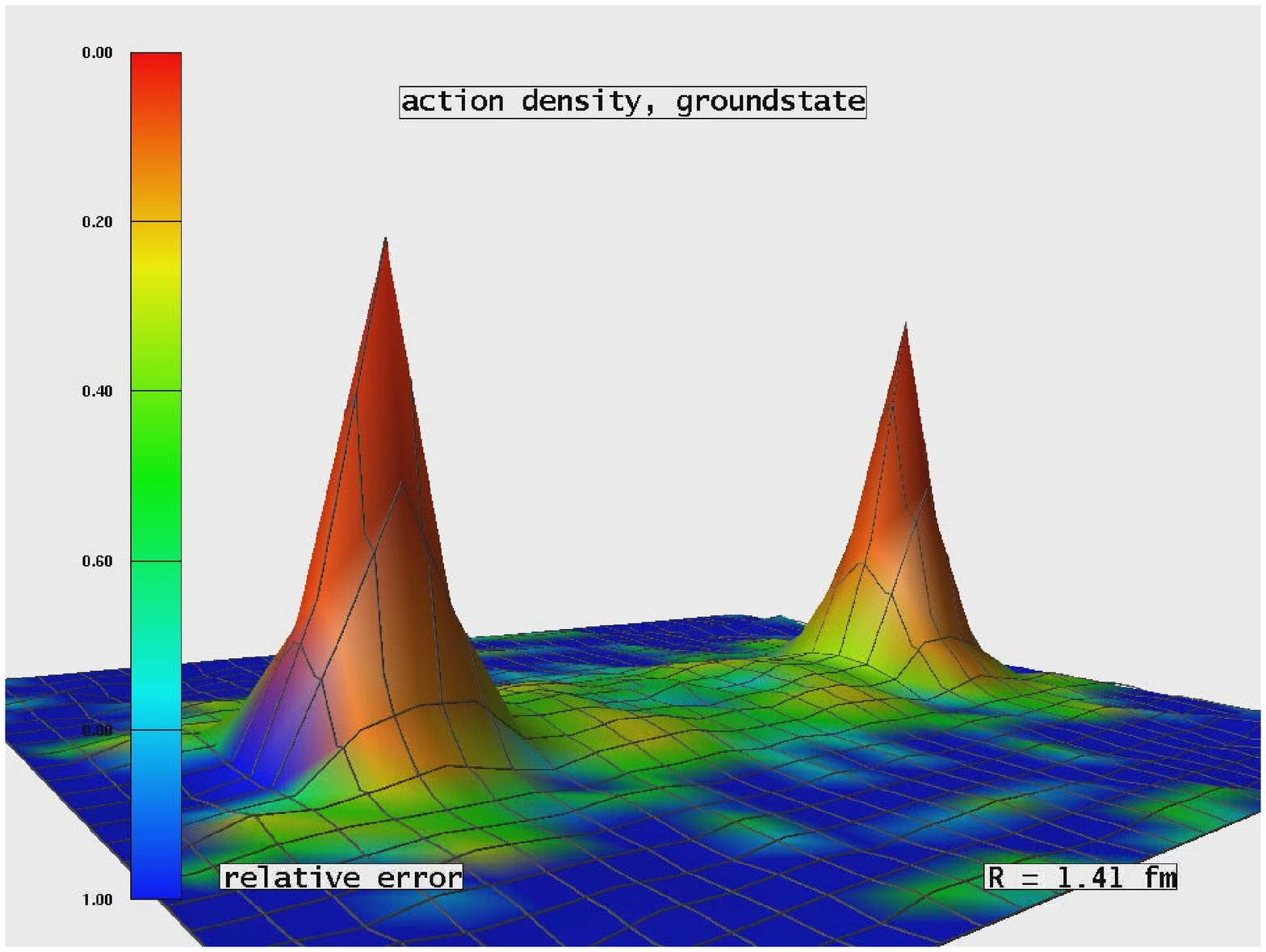,width=0.49\textwidth}\\[-1.2cm]
\caption{Ground state action density distribution at
$r> r_c$.}
\label{fig:frame2}
\end{figure*}

Diagrammatically,
$g(r)$ can be represented as,
\begin{equation}
g(r)=\frac{n_f\wwb}{\left(\www\times n_f^2\wbbd\right)^{1/2}}
\propto \sqrt{\frac{n_f}{N_c}},
\end{equation}
where $N_c$ is the number of colours:
the large $N_c\gg n_f$ expectation
for the minimal energy gap is, $\Delta E_c\propto\sqrt{n_f/N_c}$.
Obviously, precision studies
at different $N_c$ and $n_f$ are needed to establish the
validity range of this prediction.
The decay of the adjoint potential into two gluelumps,
where $\Delta E_c\propto 1/N_c$, also fits nicely into this context.

The quality of our density distribution data is depicted in
Figure~\ref{fig:transverse} for the ground state at
a separation slightly smaller than $r_c$, as
a function of the transverse distance $x$
from the $\overline{Q}Q$ axis. 
Due to cancellations between the magnetic and
electric components the energy density is much smaller than the action
density: for the comparison we
have multiplied the energy density data by the arbitrary factor of 29.
Note that the ratio $\sigma/\epsilon$ will diverge like $-\ln a\Lambda$
in the continuum limit. The differences between the shapes
of the energy and action density distributions are not statistically
significant. Below we only visualise the more precise action density results.

We employ several off-axis separations. Assuming
rotational symmetry about the interquark axis, each
point is labelled by two coordinates.
$x$ denotes the distance from the $\overline{Q}Q$ axis
and $y$ denotes the longitudinal distance from the centre point.
We define an interpolating rectangular
grid with perpendicular lattice spacing $a$ and the longitudinal spacing
slightly scaled, such that the static sources always lie on
integer grid coordinates. We then assign a quadratically interpolated value
to each grid point ${\mathbf z}$, obtained from points
in the neighbourhood,
$|{\mathbf z}-(x,y)|\leq \epsilon=a$.
On the axis the data points are more sparse and we relax the condition
to $\epsilon=\sqrt{3}a$ while for the singular peaks we maintain the
un-interpolated values.

In Figures~\ref{fig:frame} and \ref{fig:frame2} we display
ground state action density distributions for different
distances around $r_c$. An mpeg animation has been published
in Ref.~\cite{Prkacin:2005dc}.
The colour encodes the
relative statistical errors and the lattice mesh represents
our spatial resolution $a$.
As already evident from the mixing angle $\theta(r)$ of
Figure~\ref{fig:angle} above,
string breaking takes place within a small region
around $r_c$. All distributions are very similar to
linear superpositions of string and
broken-string states, with no non-trivial spatial dependence:
string breaking appears to resemble an
instantaneous process, without evidence of
localisation of the $q\bar{q}$ pair creation.

\section{APPLICATION TO QUARKONIUM DECAY}
We wish to relate the
static limit results to strong decay rates of
quarkonia. In the non-relativistic limit of heavy quarks,
potential ``models'' provide us with the natural
framework for such studies. In fact at short distances,
$r\ll 1/\Lambda$, potential ``models'' can be derived as
an effective field theory, potential NRQCD,
from QCD~\cite{Brambilla:2004jw}. One can in principle add a
$B\overline{B}$ sector, as well as transition terms between the
two sectors, to the $\overline{Q}Q$ pNRQCD Lagrangian. Strong
decays would then be a straightforward non-perturbative
generalisation
of the standard multipole treatment of radiative transitions
in QED. Unfortunately, transitions such as
$\Upsilon(4S)\rightarrow B \overline{B}$ can hardly be classed
as ``short distance'' physics. So, some modelling is required instead.

The natural starting point again is a two channel
potential model which might still have some validity
beyond the short distance regime:
\begin{equation}
H\psi({\mathbf r})=E\psi({\mathbf r})
\end{equation}
with
\begin{equation}
H=\left(\begin{array}{cc}\frac{1}{m_Q}&0\\
0&\frac{1}{m_B}\end{array}\right){\mathbf p}^2+V(r).
\end{equation}
$m_Q$ denotes the heavy quark mass and $m_B$ is the mass
of a $B$ meson. The wavefunction has two components,
\begin{equation}
\psi({\mathbf r})=\left(\begin{array}{c}\psi_{\overline{Q}Q}({\mathbf r})\\
\psi_{B\overline{B}}({\mathbf r})\end{array}\right)
\end{equation}
and the potential is given by,
\begin{eqnarray}
V(r)&=&O^{\dagger}\left(\begin{array}{cc}E_1(r)&0\\0&E_2(r)\end{array}\right)O\\
&=&\left(\begin{array}{cc}E_Q(r)&g(r)\\g(r)&E_B(r)\end{array}\right),
\end{eqnarray}
where we have normalized the zero point energy to $2m_B$ and
\begin{equation}
O=\left(\begin{array}{cc}\cos\theta&\sin\theta\\-\sin\theta&\cos\theta\end{array}\right)
\end{equation}
rotates our Fock basis $\{|Q\rangle,|B\rangle\}$ into the eigenbasis
$\{|1\rangle,|2\rangle\}$.
We neglect $B$-$\overline{B}$ interactions and set $E_B(r)=0$.
We further adjust the difference
$m_{\Upsilon(4S)}-2m_B$ to the experimental value which affects
our phase space factor.
We then
follow Ref.~\cite{Drummond:1998eh} and
calculate the decay rate by multiplying phase space
with the overlap integral
between the $\Upsilon(4S)$ wave function, the interaction term
$g(r)$ and the $B\overline{B}$ continuum.
In doing so, we assume that the interaction does not contain any spatial
distribution but only depends on the
distance $r$. This instantaneous picture
is supported by our action density measurements above.

\begin{figure}[th]
\epsfig{file=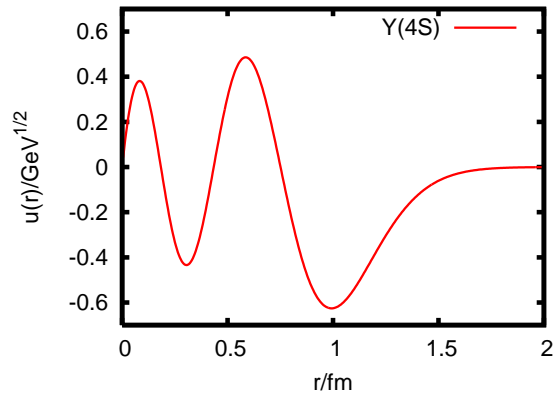,width=0.48\textwidth}\\[-1.2cm]
\caption {The radial $\Upsilon(4S)$ wave function.}
\label{fig:wave}
\end{figure}
In Figure~\ref{fig:wave} we display our radial $\Upsilon(4S)$ wave function
$u(r)=\sqrt{4\pi}\,r\,\psi_{\overline{Q}Q,400}(r)$. The decay rate depends
rather sensitively on the positions of the nodes.
We obtain a preliminary result $\Gamma[\Upsilon(4S)\rightarrow B\overline{B}]
\approx 5$~MeV, which is about half the experimental value.
This appears very reasonable, given the crudeness of
the model and the fact that the gap $\Delta E_c$ will increase
with lighter, more realistic sea quark masses. We are studying the situation
and systematics in more detail.

\section{CONCLUSIONS}
We were able to resolve the string breaking problem in $n_f=2$ QCD,
at one value of the lattice spacing $a^{-1}\approx 2.37$~GeV and
of the sea quark mass, $m\lesssim m_s$. It was also possible
to study the dynamics of string breaking in detail and to resolve
spatial colour field distributions. The breaking of the string
appears to be an instantaneous process, with de-localized light
quark pair creation. While a direct lattice study of strong
decay rates such
as $\Upsilon(4S)\rightarrow B\overline{B}$ or $\Psi(3S)\rightarrow
\overline{D}D$ is at present virtually impossible, investigations
in the static limit can help constraining models.
Studying the energy between pairs of
static-light mesons can also be viewed as
a milestone with respect to
future calculations of $\Lambda_Q\Lambda_Q$ forces,
which are related to nucleon-nucleon interactions~\cite{Arndt:2003vx}.

\section*{ACKNOWLEDGMENTS}
The computations have been performed on the IBM
Regatta p690+ (Jump) of ZAM at FZ-J\"ulich and on the ALiCE cluster computer
of Wuppertal University. This work is supported by the
EC Hadron Physics I3 Contract RII3-CT-2004-506078,
by the Deutsche Forschungsgemeinschaft and by PPARC.

\end{document}